\documentclass[epsfig,12pt]{article}
\usepackage{graphicx}
\usepackage{amsmath}

\input epsf.sty
\newtheorem{theorem}{Theorem}
\newtheorem{acknowledgement}[theorem]{Acknowledgement}

\input{tcilatex}
\makeatletter \@addtoreset{equation}{section}

\def\be{\begin{equation}}
\def\ee{\end{equation}}
\def\bea{\begin{eqnarray}}
\def\eea{\end{eqnarray}}

\newcommand{\nc}{\newcommand}
\nc{\al}{\alpha} \nc{\bib}{\bibitem} \nc{\la}{\lambda}
\nc{\C}{\mbox{\hspace{1.24mm}\rule{0.2mm}{2.5mm}\hspace{-2.7mm}
C}} \nc{\R}{\mbox{\hspace{.04mm}\rule{0.2mm}{2.8mm}\hspace{-1.5mm}
R}}
\begin{document}

\title{%
\rightline{\mbox {\normalsize
{Lab/UFR-HEP0704/GNPHE/0704}}\bigskip}\textbf{\ On
Brane Inflation Potentials  and Black Hole  Attractors }}
\author{Adil  Belhaj$^1$\thanks{belhaj@unizar.es}, Pablo  Diaz$^1$\thanks{pdiaz@unizar.es}, Mohamed  Naciri$^{2,3}$\thanks{mhnaciri@gmail.com}, Antonio  Segui$^{1}$\thanks{segui@unizar.es}\\
{\small 1. Departamento de Fisica Teorica, Universidad de Zaragoza, 50009-Zaragoza, Spain.}\\
{\small 2. Lab/UFR-Physique des Hautes Energies, Facult\'{e} des
Sciences (FS), Rabat, Morocco.}\\
{\small 3. Groupement National de Physique des Hautes Energies, Si\`{e}ge focal: FS Rabat, Morocco.}
} \maketitle

\begin{abstract}
\bigskip
We propose a new potential in brane inflation theory, which is given by 
the arctangent  of the   square of the scalar field.
Then we perform an explicit computation for inflationary quantities. This
potential has many nice features. In the small field approximation, it
reproduces the chaotic and MSSM potentials. It allows one, in the large
field approximation, to implement the attractor mechanism for bulk black
holes  where the geometry on the brane is  de Sitter. In particular, we show, up 
to some assumptions, that the Friedman equation can be reinterpreted as a
Schwarzschild black hole attractor equation for its  mass parameter.

\textbf{Key words}: Inflation, Braneworld model, Attractor mechanism, Black
hole.
\end{abstract}


\newpage

\newpage

\section{Introduction}

Recently, an increasing interest has been devoted to the study of inflation
in connection with string theory \cite{Br,B,C,L1,PRZ,KKLMMT},  black hole \cite
{Ru,A} and brane physics \cite{RS1,RS2}.  In this regard, one of the famous
work is the Randall-Sundrum (RS) model which was shown to be classified into
two types: A model with two branes with opposite tensions  and another in which there is  
only one brane with positive tension. It may be thought of as
sending the negative tension brane off to infinity.

The scalar potential shape turns out to be essential in  inflationary models
\cite{R}. The well known examples are  the chaotic inflation potential
dealt with in \cite{MWBH1,PZ1} and  the minimal supersymmetric standard model
(MSSM) inflation studied recently in \cite{AGBEM,L}. Besides these examples,
there are several other models which have been much discussed in the  literature.
In particular,  one has  the exponential and the inverse power-law
potentials \cite{L2}  which may be used in the study of quintessence in
brane and tachyonic inflation.

One of the aims of the present work is to contribute to this program by
proposing a new potential in the  brane inflation scenario. This potential
is given by $V(\phi )=\lambda f(\frac{\phi }{\nu })$, where $f(\frac{\phi }{%
\nu })=arctg(\phi ^{2})$ and $\lambda $ is a mass parameter specified
later on. In this study $\nu $ is fixed to one. An objective in this paper
is to give an explicit computation for inflationary quantities corresponding
to this potential. Then we make contact with observational results. We will
see that, for a large range of  values of $\lambda $, one can get results in
agreement with the observations.

On the other hand, the potential that we propose here has interesting
features. In the very small field regime,  it  behaves as the
chaotic  potential. In this way, $\lambda $ can be  related to the 
inflaton mass. \ If we expand \ our potential \ at order 10, it can be identified
with the \ one involved in \ the MSSM inflation. We will see that, a trivial
identification of these potentials puts strong constraints on the MSSM
parameters. Finally, one of the nice features of our potential is that, in
the large field approximation, it allows one to implement \ the attractor
mechanism in the presence of a black hole in the bulk. \ \ For  a
Schwarzschid black hole, we show, up to some assumptions, that its mass is
proportional to $\lambda $ defining the asymptotic value of the potential.
In this way, the Friedman equation can be interpreted as an attractor
equation involving the mass parameter of  a bulk  black hole  when the geometry
on the brane  is  de Sitter one. We consider  the presence of
two effective cosmological constants; one drives the inflationary era and the other
accounts for the  current observations. 
The attractor mechanism  with a bulk black hole is 
the responsible for the cancellation of the inflationary cosmological constant, 
as observed today. 

After a brief summary of brane inflation in section 2, we propose the new
potential, and then we discuss its inflationary aspects in section 3. The
idea is to give a detailed computation for some inflation quantities and
make contact with the observations. Section 4 concerns  with a  connection between  the 
black hole  in the bulk and the scale of the potential.  
In particular, we interpret the
Friedman equation as an attractor equation for the Schwarzschid black hole
in the bulk. In this sense, we find that the mass of the black hole can be
related to the asymptotic value of the potential. A conclusion is
presented in section 5.

\section{ Overview on brane inflation}
In this section, we shortly  review some basic facts on $3$-brane cosmology in
five-dimensional space-time \cite{MWBH1}. Related works can   found in \cite{RS1,RS2,BV,BDEL,MWBH2,M,FTW,PZ2,Li}.   We assume that the
universe is filled with a perfect fluid with energy density $\rho (t)$ and pressure $p(t)$. In the framework of the flat Friedmann-Robertson-Walker model, with a scale
factor $a(t)$, the Friedman equation  is 
\begin{equation}
H^{2}=\left( \frac{\dot{a}}{a}\right) ^{2}=\frac{8\pi }{3M_{4}^{2}}\rho %
\left[ 1+\frac{\rho }{2T}\right] +\frac{\Lambda _{4}}{3}+\frac{\mu }{a^{4}},
\label{friedman1}
\end{equation}
describing the time evolution of $a(t)$.  Here $H(t)=\frac{\dot{a}}{a}$ defines
the Hubble parameter  and the first term  of (\ref{friedman1}) is 
the responsible for inflation.  $M_{4}$  is the $4$-dimensional  Planck scale    which is related to the  $5$-dimensional one  $M_{5}$  by 
$M_{4}=\sqrt{\frac{3}{4\pi }}\left( \frac{M_{5}^{2}}{\sqrt{T}}\right) \ M_{5}$
where $T$ is the $3$-brane tension.
$ \Lambda _{4}$  is the current cosmological constant.   $\mu $ is related to the mass of the black hole in the bulk   by $
\mu=M_{BH}G_N^{(5)} $ where  $G_N^{(5)}$ is the Newton constant in five dimensions
\cite{BV}.
Consider now an inflationary theory driven by a scalar field $\phi $. The
dynamics can be described by a perfect fluid with a time dependent energy
density  $ \rho =\frac{1}{2}\dot{\phi}^{2}+V(\phi )$ and pressure  
$ p =\frac{1}{2}\dot{\phi}^{2}-V(\phi )$.  $V(\phi )$ is the dominant energy contribution to inflation. 
The scalar field satisfies the Klein-Gordon equation, namely,
\begin{equation}
\ddot{\phi}+3H\dot{\phi}+V^{\prime }\left( \phi \right) =0,\;\;\;\;\;  V^{\prime }=\frac{dV}{d\phi }.\label{kge1}
\end{equation}%
 The   dynamics of  inflation requires  that
the scalar field moves away from  the false vacuum and slowly rolls down to
the minimum of its effective potential \cite{L2}.  
The  slow-roll condition is 
characterized by two parameters
\begin{eqnarray}
\epsilon =\frac{M_{4}^{2}}{4\pi }\left( \frac{%
V^{^{\prime }}}{V}\right) ^{2}\left[ \frac{T(T+V)}{\left( 2T+V\right) ^{2}}%
\right], 
  \;\;\;\;\;\;
\eta  =\frac{M_{4}^{2}}{4\pi }\left(
\frac{V^{\prime \prime }}{V}\right) ^{2}\left[ \frac{T}{2T+V}\right] .
\notag  \label{srp}
\end{eqnarray}
 Inflation ends  when any of the  two  parameters equals one. In the slow-roll approximation,  these parameters are very small,
namely  max$\left\{ \epsilon ,\left\vert \eta \right\vert \right\} <<1$. In this case, it is easy to compute the number of e-foldings
between the beginning and the end of inflation, given by $%
N=\int\limits_{t_{i}}^{t_{e}}Hdt$. If $\phi _{i}$ and $\phi _{e}$ are
the values of the scalar field at the beginning and at the end of inflation respectively,
then $N$ takes the following  form
\begin{equation}
N=-8\frac{\pi }{M_{4}^{2}}\int\limits_{\phi _{i}}^{\phi _{e}}\frac{V}{%
V^{\prime }}\left[ 1+\frac{V\left( \phi \right) }{2T}\right] d\phi .
\label{n1}
\end{equation}%
The inflationary model can be tested by computing the spectrum of
perturbations produced by quantum fluctuations of fields around their
homogeneous background values. In the large brane tension limit, the
scalar amplitude $A_{s}^{2}$ of density perturbation is given by
\begin{equation}
A_{s}^{2}\sim \frac{512\pi }{75M_{4}^{6}}\frac{V^{3}}{V^{^{\prime }2}},
\label{cpert}
\end{equation}%
and the spectral index  reads as 
\begin{equation}
n_{s}=1-6\varepsilon +2\eta .  \label{as}
\end{equation}

The above brane-world formalism has been applied for inflationary 
models  involving many potential forms. In the present work, we shall
apply this formalism for a new potential and give a detailed computation for
some  inflation quantities. Then we make contact with black hole  attractor.

\section{A new inflation potential}

In this section we propose a new potential for inflation.
Recall that a generic single  scalar field potential, which can be characterized by
two independent parameters, has the following form
\begin{equation}
V(\phi )=\lambda f(\frac{\phi }{\nu })  \label{potential}
\end{equation}%
where $\lambda $ corresponds to the vacuum energy density and $\nu $
corresponds to changes in the field value $\Delta \phi $ during inflation,
which will be fixed to one \cite{R}. Different models can be obtained by
taking different choices for the function $f$. The most famous example, which
has been intensively studied,   is the chaotic inflation and its phenomenological 
hybrid extension \cite{L1}. It has the
following  form
\begin{equation}
V\left( \phi \right) =\frac{1}{2}m^{2}\phi ^{2},  \label{chaotic}
\end{equation}%
where $m$ is the mass of the inflaton. Besides this example, there are
several other models which have been much discussed in  the literature \cite{L1}. As
mentioned in the  introduction, they involve, among others, the
exponential potential $V(\phi )=V_{0}exp(-\beta \phi )$ and the inverse
power-law potential $V(\phi )=\frac{\mu ^{\alpha +4}}{\phi ^{\alpha }}$.

The brane inflation potential that we propose to study is
\begin{equation}
V\left( \phi \right) =\lambda f(\frac{\phi }{\nu })=\lambda arctg(\phi ^{2}).
\label{f}
\end{equation}%
This potential has many nice features, we will just quote some of them. 
For very small values of the field, the potential behaves as the  chaotic one. 
Indeed, when $\phi ^{2}$ is very close to zero, at second order, one
gets a chaotic inflation scenario with 
\begin{equation}
V\left( \phi \right) =\lambda \phi ^{2},
\end{equation}%
where now $\lambda $ is related to the mass of the inflaton as
\begin{equation}
\lambda =\frac{1}{2}m^{2}.
\end{equation}%
On the other hand, it has been proposed in \cite{L,BDL} a potential,
corresponding to soft supersymmetry breaking, with the form
\begin{equation}
V\left( \phi \right) =\frac{1}{2}m^{2}\phi ^{2}-A\frac{\alpha _{p}\phi ^{p}}{%
pM_{p}^{p-3}}+\alpha _{p}^{2}\frac{\phi ^{2(p-1)}}{pM_{p}^{2(p-3)}},
\label{mssm1}
\end{equation}%
to study the MSSM inflation. In particular, it has been suggested in \cite{AGBEM}
that we are dealing with one of the flat directions of the MSSM. In particular, for 
$p=6$ we have
\begin{equation}
V\left( \phi \right) =\frac{1}{2}m^{2}\phi ^{2}-A\frac{\alpha _{6}\phi ^{6}}{%
6M_{6}^{3}}+\alpha _{6}^{2}\frac{\phi ^{10}}{6M_{6}^{6}}.  \label{mssm2}
\end{equation}%
This is a new type of inflation model which might work with any flat
direction generating an A-term. More details can be found in \cite{AGBEM,L,BDL}.
It is easy to see that (\ref{mssm2})  can be rederived from (\ref{f}).  In
the small filed approximation, the Taylor expansion at order 10 of the
function (\ref{f}) gives the following potential
\begin{equation}
V\left( \phi \right) =\lambda (\phi ^{2}-\frac{1}{3}\phi ^{6}+\frac{1}{5}%
\phi ^{10})
\end{equation}%
which has a form  similar to  the MSSM potential given in (\ref{mssm2}). A
trivial identification constrains the MSSM parameters:
\begin{equation}
\frac{1}{2}m^{2}=\frac{A\alpha _{6}}{2M_{6}^{3}}=\frac{5\alpha _{6}^{2}}{%
6M_{6}^{6}}.
\end{equation}%
Lastly  we   mention that  the computation  of
 some inflation  quantities   is  in agreement with the observable results \cite{MWBH1}. Let us
discuss this feature in detail. Within the slow-roll regime, we shall 
compute the spectral index which is a essential  quantity in
inflation theory. To obtain it, we need to calculate the value of
the scalar field $\phi _{i}$ at the beginning of inflation. A way
to get this value is by using the formula for the number of e-foldings  and then make appropriate  assumptions. Imposing that the scalar field $\phi $ is very large, $N$, 
in our model, can be written as
\begin{equation}
N\simeq 4\frac{\pi ^{2}}{M_{4}^{2}}(1+\frac{\lambda \pi }{4T})\int_{\phi
_{e}}^{\phi _{i}}\frac{1+\phi ^{4}}{\phi }d\phi .  \label{n2}
\end{equation}%
Identifying $\phi _{i}$ with $\phi _{cobe}$ and getting $\phi _{e}$ from the
end of inflation, we can find an expression for $N$. Performing the
integral, within the slow roll regime and at high energy physics $\ (V\gg T)$
we get
\begin{equation}
N\simeq \frac{\lambda \pi 3}{TM_{4}^{2}}\left[ \ln \frac{\phi _{cobe}}{\phi
_{end}}+\frac{1}{4}\left( \phi _{cobe}^{4}-\phi _{end}^{4}\right) \right] .
\label{ef}
\end{equation}%
Note that, after neglecting the first term, this equation is similar to
the usual chaotic inflation one, which is given by
\begin{equation}
N\simeq m^{2}\left( \phi _{cobe}^{4}-\phi _{end}^{4}\right) .  \label{nef}
\end{equation}%
Setting $\varepsilon_{end}\simeq 1$, \ which defines the end of inflation,
and taking \ the limit $\lambda \pi >>2T$, we obtain
\begin{equation}
\phi _{end}^{4}\simeq \frac{3}{32\lambda \pi ^{5}}M_{5}^{6}.  \label{phiend}
\end{equation}%
In order to estimate the value of \ $\phi _{cobe}$, we assume that the
second term on the right \ of (\ref{ef}) dominates. Then, we can deduce the
expression of $\phi _{cobe}$ in terms of the number of e-foldings  $N$
\begin{equation}
\phi_{cobe}^{4}\simeq \frac{3N}{\lambda \pi ^{4}}M_{5}^{6}.  \label{phicobe2}
\end{equation}%
As we show below, to test our inflation model we have to compute the
spectrum of perturbations. Indeed, a simple calculation from \ (\ref{cpert})
reveals \ that the spectrum of perturbations takes the following form
\begin{equation}
A_{s}^{2}\simeq \frac{\pi ^{7}\lambda ^{6}}{300T^{3}M_{4}^{6}}\phi
_{cobe}^{6}.  \label{as1}
\end{equation}

The scaling relation between the tension and the mass of the inflaton which is consistent with the observations is obtained from (\ref{phicobe2}) and  (\ref{as1}) and reads 
\begin{equation}
A_s \sim \frac{\lambda^3 }{T^{3/2}} \frac{\phi_{cobe}^3}{M_4^3}.
\end{equation}
The analogous result was obtained in  \cite{MWBH1} 
\begin{equation}
A_s \sim \frac{\lambda^2 }{T^{3/2}} \frac{\phi_{cobe}^5}{M_4^3}.
\end{equation}
 and  plotted in figure 1.  We can now make use of numerical results. We assume that the number of
e-foldings before the end of inflation, at which observable perturbations are
generated, corresponds to $\ N\simeq 55$ \ and \ we take  $A_{s}=2.10^{-5}$.
Varying $\phi _{cobe}$ in terms of $\lambda $ and taking different values of
$M_{5}<10^{17}Gev$, we can see that $\phi _{cobe}$ is always less than $M_{4}
$ which is consistent with the observations.

Using the above results and the same method used in the literature \cite{MWBH1}, one can easily obtain the spectral index for our
model. A straightforward computation from \ (\ref{as}) shows that
\begin{equation}
1-n_{s}\simeq \frac{\pi }{N\lambda }.  \label{spec2}
\end{equation}%
Fixing $N$, the spectral index will depend only on the parameter $\lambda $.
When $\lambda \rightarrow \infty $, the spectral
index is driven towards the Harrison-Zeldovich spectrum, $%
n_{s}= 1.$

As we have seen at the beginning of this section, in the small field regime,
$\lambda $ is related to the mass of the inflaton in the chaotic scenario. This
can be seen directly from the formula  for  the number   of e-foldings  (\ref{nef}).
One might ask the following question: Is there any interpretation of $%
\lambda $ in the large field  limit?. This question will be addressed in the
next section.

\section{ Large field approximation and  the attractor mechanism in
inflation}

In this section we consider the large field approximation for the potential
proposed in the  previous section and  then make contact with the  attractor
mechanism  in the presence of  a  black hole in the bulk. Motivated by the
result of the attractor mechanism given in \cite{KSS,AGM}, we will interpret the  Friedman equation (\ref%
{friedman1}) as an attractor equation involving the mass of the black hole.
To do this, let us consider nonzero values for $\Lambda _{4}$ and $\mu $.
Recall that the parameter $\mu $  is  the mass of the black hole in the bulk. 
For simplicity, we consider the case of a Schwarzschild black hole which 
is characterized only by its mass.

Let us take the  large field approximation. In this regime, one has
\begin{equation}
V(\phi )\sim \lambda .  \label{lfa}
\end{equation}%
In the vanishing limit $\mu =0$, it follows, from the Friedman equation,
that $\ H(t)$ should be constant, defining a de Sitter \ geometry. In this
case  the Universe expands exponentially and so the scale factor is given by
\begin{equation}
a(t)=a_{0}\exp (Ht),  \label{scale}
\end{equation}%
where $a_{0}$ is a free constant parameter, which can be fixed to one. By
integration (\ref{kge1}), one can derive the time evolution of  the scalar
field  which is consistent with  (\ref{scale}) . In
the above approximation, (\ref{kge1}) can be simplified as
\begin{equation}
\ddot{\phi}+3H\dot{\phi}=0.  \label{kge2}
\end{equation}%
Using (\ref{scale}), the last equation is solved by
\begin{equation}
\phi (t)=\xi +\exp (-3Ht),  \label{evo1}
\end{equation}%
where $\xi $ \ is a constant field. At this point, there are some comments. 
The first one is that, in the case of $\xi =0$,
the field solution becomes
\begin{equation}
\phi (t)=\exp (-3Ht).  \label{evo1}
\end{equation}%
and goes  rapidly to zero so that the  potential   tends  to the
chaotic limit. The second one is that, we may interpret $\xi $ as the value
of the scalar field when $t$ goes to infinity and so we can write
\begin{equation}
\phi (t)=\phi _{\infty }+\phi _{0}\exp (-3Ht).  \label{evo12}
\end{equation}%
The large field approximation requires that $\phi _{\infty }$ should be very large. In
this way, the scalar field is almost constant and tends to a very large
value as $t$ tends to infinity.

Take now a generic non zero value  for $\mu $. In this way $H$ is no longer
constant. However, a constant value for $H$ requires a particular form for
the mass of the black hole in the bulk. A simple inspection shows  that $\mu
$ should take the following form
\begin{equation}
\mu \sim a^{4}.  \label{masbbh1}
\end{equation}%
With this condition at hand we will interpret the \ Friedman equation as an
attractor equation. To do so, we will restrict ourself to a de Sitter \
geometry  corresponding to a specific constraint  on  $H$\footnote{%
Note that this equation may be interpreted as the 4-dimensional Einstein
equations on the brane in the presence of a positive cosmological constant.}
\begin{equation}
3H^{2}=\Lambda _{4}. \label{lambda}
\end{equation}%
For a large value of the brane tension, we can show that
\begin{equation}
\mu (t)\sim-\lambda e^{4Ht}.  \label{masbbh1}
\end{equation}%
In the case where $H$ takes small values, that is in the limit $Ht<<1$, the
mass of the black hole becomes
\begin{equation}
\mu (t)=\mu \sim -\lambda.  \label{masbbh2}
\end{equation}
 This equation relates the mass of the black hole in the bulk with the asymptotic value of the
potential. If we forget, for a while, about the negative sign,   it  can be
interpreted as an attractor equation for a Schwarzschild black hole. Note that (\ref{masbbh2}) is slightly different from the one used in 
the attractor mechanism  of $N=2$ supersymmetric  extremal black hole embedded in string theory  compactified on Calabi-Yau threefolds \cite{KSS}.
In our case we obtain a relation involving the mass of the black hole instead of the mass square as usually appears in the $N=2$ attractor equations. 
The mass parameter,  in our case,  seems to behave as the charge in the attractor mechanism, although it is fixed at infinity and not at the horizon. We believe that this difference is due to the absence of supersymmetry in this study.
It has been  discovered a sort of susy (pseudo-susy) in a relation between domain wall and cosmological solutions \cite{ST}; the deviation of our analysis from the standard result might be related to this fact.
Recently a relation between the wrapped D3 and D5-branes on cycles  of the  resolved  conifold  and inflation has been reported in  \cite{BLS,G}. 
Our model can be understood by the supersymmetric ones when susy is broken by the presence of  D$3$-D$5$  brane  system and fluxes; a sort  of pseudo-susy would  be behind the attractor mechanism we comment before. In this way, the above scalar field appearing in our potential could be identified  with   a scalar mode of the  $R$-$R$   $B$-field  on  the two-cycle of the small resolution of the  conifold.   In connection with  non supersymetric Calabi-Yau  black hole attractor \cite{KM}, it should be interesting to determine the  entropy function  using flux compactification	 on  the resolved  conifold.  We hope some new results in this direction  can be presented in the future \cite{work}.

As usually the instability corresponding to negative mass can be solved by
introducing a brane with negative tension. Indeed, using the Friedman equation
and (\ref{lambda}), we have
\begin{equation}
\mu =-\frac{2\pi ^{2}}{3M_{4}^{2}}\lambda a_0^4\left[ 1+\frac{\pi \lambda }{2T}%
\right] .  \label{masbbh4}
\end{equation}%
This equation indicates that the mass of the  black hole in the bulk is
related to the brane tension and the value of the potential at infinity. In
the  high energy limit, $\pi \lambda >>2|T|$, (\ref{masbbh4}) gets reduced to
\begin{equation}
\mu =\frac{\pi ^{3}}{3|T|M_{4}^{2}}\lambda ^{2}a_0^4.  \label{masbbh5}
\end{equation}
From  this relation it is clear that in order to interpret it as an attractor equation $\mu \simeq \frac{\pi }{2}\lambda =V(\phi )|_{\phi \rightarrow \infty }$, a constraint on the brane  tension is required. This is given by 
\begin{equation}
|T|=\frac{\pi ^{2}\lambda }{3M_{4}^{2}}a_0^4.
\end{equation}
 We interpret this  result in the following way. The  black hole  in the
bulk is placed near the second  brane with \ negative tension $(T<0)$. 
The radiation of this black hole  contributes to the
Friedman equation (\ref{friedman1}) with the term $\frac{\mu }{a^{4}}.$

\section{Conclusion}

\qquad  In this paper, we have proposed a new potential in the brane
inflation scenario. After computing the main inflation parameters and check that they agree with the observational results we have implemented the attractor mechanism using a black hole in the bulk in a particular way. 
The discussion of the fourth section is easily extended to more general potentials. It is only necessary a constant positive value of the potential at infinity. This kind of potentials are, in general, consistent with the slow roll conditions. Our potential has an extra bonus.
In the small field regime, it reproduces
the chaotic model as well as the MSSM inflation potential.

 The result may be summarized
as follows:

(1) First, we have achieved a quantitative study and computed inflationary
quantities. In order to make contact with the observational  data, several
bounds can be imposed to our potential so that the results are in agreement.

(2) In the  large field approximation, we have made contact with the black
hole in the bulk. For a Schwarzschild black hole solution, we have shown that
its mass is proportional to $\lambda $ defining the asymptotic value of the
potential. This has been done with  the help of the Friedman equation, which
may be reinterpreted as an attractor equation in the de Sitter background.

Our work opens up for further studies.  In connection with inflation in string theory, one may  consider models with several scalar fields $\phi_i$. These fields  can identified   with the $R$-$R$  $B$-field on $2$-cycles of the Calabi-Yau manifolds. In this way, the total potential takes the following form
\begin{equation}
V(\phi_i)=\sum_i\lambda_i\; arctg(\phi_i ^{2}),\;\;\;\;\; i=1,...,h^{1,1}
\end{equation} where  $h^{1,1}$  is the  number of the Kahler deformations of the Calabi-Yau  manifold. 
  Another  natural extension of the present
results includes other type of black hole solutions. 
We think that the relation  between the inflationary
potential and \ the attractor mechanism in the Calabi-Yau   black hole physics deserves a better understanding. We hope
to report elsewhere on these open questions.

\begin{acknowledgement}
\qquad\ \newline
We thank  E. H. Saidi   for discussions and collaborations related to this work.
AB and MN would like to thank  R. Ahl Laamara  and  J. Houda for  discussions.  AB  thanks  S.  Montanez for discussions and hospitality. This work has been supported by MCYT ( Spain) under grant FPA
2003-02948. MN is supported by the program Protars III D12/25, CNRST (
Morocco).
\end{acknowledgement}

\end{document}